\documentstyle[twoside]{article}
\def\Td#1#2#3#4#5{{\thispagestyle{empty}
\protect\headheight0pt\protect\headsep0pt\protect\vspace*{-2.2cm}
{\flushleft\parbox{143mm}{\footnotesize Commun.\ Theor.\ Phys.~(Beijing,
China) {\bf #1}{~(#2)~}{pp~#3}\\[-0.7mm]
\copyright\hspace*{3.5pt} International Academic Publishers\hfill Vol.~{#4},
No.~{#5}}\\[-1.4mm]
\begin{table}[h]\hfill\null\hfill\hrule\vskip.4mm\hrule\end{table} }}}
\def\wen#1{$^{[#1]}$}
\def\ssc#1{\scriptscriptstyle{#1}}
\def\dsize{\displaystyle}
\def\tr{\mathop{\rm \,tr\,}\nolimits}
\def\no#1{\rlap{\protect\rule[-0.25 true cm]{\textwidth}{0.03 true cm}}%
No.~{#1}\hfill}
\def\vo#1{\hfill{\ignorespaces Vol.~{#1}%
\llap{\protect\rule[-0.25 true cm]{\textwidth}{0.03 true cm}}}}
\def\ld{\protect\footnotesize}
\def\rd{\protect\footnotesize}
\def\title#1{\begin{flushleft}
\Large\bf\protect\baselineskip=17pt #1\end{flushleft}}
\def\author#1{\leftskip=1cm\noindent\begin{minipage}{133mm}
\normalsize\hspace*{-4.5pt}#1\end{minipage}\par\vglue4pt}

\def\Address#1{\leftskip=1cm\noindent\hspace*{-7.0pt}
\begin{minipage}{133mm}\protect\small\baselineskip=10pt#1\end{minipage}\par\vglue2pt}
\def\date#1{\leftskip=1cm\vglue4pt\noindent\begin{minipage}{133mm}
\normalsize\hspace*{-4.5pt}#1\end{minipage}\par\vglue4pt}
\def\abstract#1#2{\leftskip=1cm\vglue4pt\noindent\hspace*{-4.5pt}\begin{minipage}{133mm}
\small\sl\baselineskip=11pt{\bf Abstract} #1\par
\pacs{\normalsize#2}\end{minipage}\par\vglue2pt\leftskip=0cm}

\def\pacs#1{\protect\vglue6pt%
\noindent\begin{minipage}{133mm}{\bf PACS numbers: }\rm #1\end{minipage}%
}
\def\key#1{\leftskip=1cm\vglue0pt\noindent\hspace*{-4.5pt}\protect
\begin{minipage}{133mm}\begin{minipage}[t]{21mm}{\bf Key words:}\end{minipage}
\hfill\begin{minipage}[t]{111mm}\rm\baselineskip=12pt #1\end{minipage}
\end{minipage}\par\leftskip=0cm}

\def\yskip{\vskip5pt plus 2pt minus 1pt}
\def\sect#1{\medskip{\noindent #1\par}\yskip}
\makeatletter
\def\d{\hspace*{.8pt}\mathstrut{\rm d}\hspace*{.6pt}}
\def\e{\mathstrut\,{\rm e\/}\hspace*{.6pt}}
\def\i{\hspace*{.6pt}\mathstrut{\rm i}\hspace*{.6pt}}
\def\noa#1{\noalign{\vskip#1pt}}

\def\im{\,\mathop{\rm Im\,}\nolimits}
\def\re{\,\mathop{\rm Re\,}\nolimits}
\if@twoside \def\ps@headings{\let\@mkboth\markboth
\def\@oddfoot{}\def\@evenfoot{}\def\@evenhead{\rm\footnotesize \thepage\hfil
\footnotesize
\leftmark}\def\@oddhead{\hbox{}\footnotesize \rightmark \hfil
\rm\thepage}\def\sectionmark##1{\markboth {\uppercase{\ifnum \c@secnumdepth
>\z@
 \thesection\hskip 1em\relax \fi ##1}}{}}\def\subsectionmark##1{\markright
{\ifnum \c@secnumdepth >\@ne
 \thesubsection\hskip 1em\relax \fi ##1}}}
\else \def\ps@headings{\let\@mkboth\markboth
\def\@oddfoot{}\def\@evenfoot{}\def\@oddhead{\hbox {}\footnotesize \rightmark
\hfil\rm\footnotesize\thepage}\def\sectionmark##1{\markright
{\uppercase{\ifnum \c@secnumdepth
>\z@
 \thesection\hskip 1em\relax \fi ##1}}}}
\fi
\def\ps@myheadings{\let\@mkboth\@gobbletwo
\def\@oddhead{\hbox{}\footnotesize\rightmark \hfill\rm\footnotesize\thepage}
\def\@oddfoot{}
\def\@evenhead{\rm\footnotesize\thepage
 \hfill \footnotesize\leftmark\hbox{}}\def\@evenfoot{}
 \def\sectionmark##1{}\def\subsectionmark##1{}}
\makeatother

\font\bfit=cmbxti10
\def\tsize{\textstyle}
\def\bk#1{\mbox{\boldmath$#1$}}
\evensidemargin=\oddsidemargin
\begin{document}
\def\fr{\frac}   \def\la{\langle} \def\ra{\rangle}

\abovedisplayskip=3pt plus 1pt minus 1pt
\belowdisplayskip=3pt plus 1pt minus 1pt
\parskip=0pt plus.1pt minus0.1pt

\setcounter{footnote}{0}
\setcounter{page}{705}
\Td{34}{2000}{705--714}{34}{4, December 15, 2000}

\title{Triality and Quantization of Singularities in Massive
Fermion}
\vglue-5pt
\author{LIU YuFen$^*$}
\vglue-4pt
\Address{Institute of Theoretical Physics, Academia Sinica, Beijing
100080, China}
\vglue-4pt
\date{(Received June 28, 2000)}
\abstract{It is proved that fermions can acquire the mass through the
additional non-integrable exponential factor. For this propose the special
vector potential associated with the spinor field was introduced.
Such a vector potential has close relation with the triality property in Dirac
spinors and plays crucial role in the construction of massive term. It is
shown that the change in phase of a wavefunction round any closed curve
with the possibility of there being singularities in our vector potential
will lead to the law of quantization of physical constants including the
mass. The triality properties of Dirac's spinors are studied and it
leads to a double covering vector representation of Dirac spinor field.
It is proved that massive Dirac equation
in the bosonic representation is self-dual.}{11.30.-j}\key{triality,
bosonisation, mass quantization}

\footnotetext[1]{Email: liuyf@itp.ac.cn}

\baselineskip=12.6pt
\sect{\large\bf I. Introduction}
\vspace*{-0.15cm}
The concept of phase is of great practical importance in contemporary
physics. For example, the theories of superconductivity and superfluidity,
the Josephson effect, holography, masers and lasers are all fundamentally
based on various aspects of the concept of phase. In this paper we introduce
a very special non-integrable phase in massive fermion, which is determined by
positive time-like vector potential $K_\mu $. This vector potential is
completely determined by spinor field and plays crucial role in the
construction of massive term. Such the  vector potential first appeared in the
redefined wavefunction of fermion in our previous work,$^{[1]}$ where the conformal
invariance of the Dirac equation was studied. From mathematical point of
view, the existence of vector $K^\mu $ is due to the existence of triality property
in Dirac spinors.

It is proved that using vector $K^\mu $, we can generate the desired massive
Dirac equation from the massless Dirac equation by substituting the massive
spinor $\Psi $ by massless spinor $\Psi _0$,
\begin{equation} 
\Psi \,\stackrel{\rm d}{=}\,\Psi _0 \e^{\i\int (eA_\mu -mK_\mu )\d x^\mu}
 =R_0 \e^{\i\int (eA_\mu -mK_\mu )\d x^\mu }+L_0 \e^{\i\int (eA_\mu
-mK_\mu )\d x^\mu }\,,
\end{equation}
such that
\begin{equation} 
\i\gamma ^\mu \partial _\mu \Psi _0=0
\end{equation}
identifies with Dirac equation for massive fermion $\Psi $.

In his famous monopole paper ``Quantization of Singularities in
Electromagnetic Field''$^{[2]}$ Dirac emphasized that
``non-integrable phases are perfectly compatible with all the general
principles of quantum mechanics and do not in any way restrict their
physical interpretation''. He allowed for wavefunctions with non-integrable
phases and conjectured that: ``{\it The change in phase of a wavefunction
round any closed curve may be different for different wave functions by
arbitrary multiplies of} $2\pi$ {\it and is thus not sufficiently definite
to be interpreted immediately in terms of the electromagnetic field''}.
These correspond to single magnetic poles with their strength restricted
by the relation $eq/(4\pi )=(n/2)$. For Monopole Meeting in 1981, 50 years
after his first paper, Dirac has sent Abdus Salam the following message:
{\it ``I am inclined now to believe that monopoles do not exist. So many
years have gone by without any encouragement from the experimental side}''.

The existence of magnetic monopole is an open question, thus in our case
it means the change in $\oint mK_\mu \d x^\mu $ round any closed curve,
with the possibility of there being singularity in ${\rm Re}(K_\mu )$, will
lead to the law of {\bfit quantization} of physical constants including
{\bfit mass}.

To illustrate the geometrical nature of $K_\mu $, we will study triality
property in Dirac's spinors, and it leads to the double covering vector representation of
spinor field. The most interesting to physicist is that:
{\it The massive Dirac equation in bosonic representation is self-dual.}

\sect{\large\bf II. The Behavior of Massive Term}
\vspace*{-0.15cm}
Let $x^\mu \subset R^{(1,3)}$ be (real) coordinates of ordinary space-time,
$\eta _{\mu \nu }={\rm diag}\,(+1,-1,-1,-1)$.
$\Psi =[({1+\gamma _5})/
2] \Psi +[({1-\gamma _5})/2] \Psi \,\stackrel{\rm d}{=}\,R+L$ is
4-component (complex) spinor, $\gamma ^\mu $ are $4\times 4$ Dirac matrices
and $\gamma ^5=\gamma _5\,\stackrel{\d}{=}\,\i\gamma^0\gamma^1\gamma^2\gamma^3$.

\vspace*{0.15cm}
\noindent {\bfit Lemma\ \ } There exists suitable complex vector
$K^\mu =K_{-}^\mu +K_{+}^\mu $ which
associates with spinor $\Psi =R+L$ (and completely determined by it) such
that
\begin{eqnarray} 
& K_\mu \gamma ^\mu R=K_\mu ^{-}\gamma ^\mu R=L \,, &\qquad
K_\mu \gamma ^\mu L=K_\mu ^{+}\gamma ^\mu L=R\,,\\
& \bar{R}K_\mu ^{*}\gamma ^\mu =\bar{R}K_\mu ^{-*}\gamma ^\mu =
\bar{L} \,, & \qquad
\bar{L}K_\mu ^{*}\gamma ^\mu =\bar{L}K_\mu ^{+*}\gamma ^\mu =
\bar{R}\,,
\end{eqnarray}
here the Dirac conjugate is defined as $\bar{R}\,\stackrel{\rm d}{=}\,
(R^{*T})\gamma _0$ and $K_\mu ^{*}$ is complex conjugation of $K_\mu $. The
explicit forms of these vectors are
\begin{equation} 
K_\mu =K_\mu ^{+}+K_\mu ^{-}=\frac{\bar{R}\gamma _\mu R}{2\bar{R}L}
+\frac{\bar{L}\gamma _\mu L}{2\bar{L}R}  \,.
\end{equation}
They satisfy
\begin{equation} 
\tsize
K_\mu \eta ^{\mu \nu }K_\nu =1 \,, \quad K_\mu ^{\pm }\eta ^{\mu \nu }K_\nu
^{\pm }=0 \,,\quad K_\mu ^{\pm }\eta ^{\mu \nu }K_\nu ^{\mp }=\frac 12\,.
\end{equation}

\noindent {\bfit Corollary\ \ }  The massive-quadratic-two form equals
 the interaction-cubic-trilinear form
\begin{equation} 
\bar{R}L=\bar{R}K_\mu \gamma ^\mu R=\bar{L}K_\mu ^{*}\gamma
^\mu L \,,\qquad
\bar{L}R=\bar{L}K_\mu \gamma ^\mu L=\bar{R}K_\mu ^{*}\gamma
^\mu R \,.
\end{equation}
If $\Psi =R+L$, then
\begin{equation} 
m\bar{\Psi }\Psi =m\bar{\Psi }K_\mu \gamma ^\mu \Psi
\end{equation}
(It is like Higgs mechanism without Higgs field).

\vspace*{0.2cm}
\noindent {\bfit Theorem\ \ } Charged massive Dirac equation can be rewritten in the following ``uncharged
massless'' forms
\begin{eqnarray} 
&&\hphantom{=\;} \i\gamma ^\mu (\partial _\mu -\i eA_\mu )R-mL
\nonumber\\
&& = \i\gamma ^\mu (\partial _\mu -\i eA_\mu +\i mK_\mu )R
= [\i\gamma ^\mu \partial _\mu R_0]\Theta ^{-1}=0\,,
\\\noa1
&& \hphantom{=\;} \i\gamma ^\mu (\partial _\mu -\i eA_\mu )L-mR
\nonumber\\
&&= \i\gamma ^\mu (\partial _\mu -\i eA_\mu +\i mK_\mu )L
=  [\i\gamma ^\mu \partial _\mu L_0]\Theta ^{-1}=0\,,
\end{eqnarray}
where $R_0\,\stackrel{\rm d}{=}\,R\e^{-\i\int
(eA_\mu -mK_\mu )\d x^\mu }$
and $
\Theta =\e^{-\i\int (eA_\mu -mK_\mu )\d x^\mu }$. It is equivalent to
\begin{equation} 
\i\gamma ^\mu (\partial _\mu -\i eA_\mu )\Psi -m\Psi =[\i\gamma ^\mu \partial
_\mu \Psi _0]\Theta ^{-1}=0\,,
\end{equation}
here $\Psi _0\,\stackrel{\rm d}{=}\,\Psi  \e^{-\i\int (eA_\mu
-mK_\mu )\d x^\mu }$ satisfies the massless Dirac equation.

It is known that a charged particle must be massive. Thus in our model the
mass $m$ was introduced in the same way as the charge $e$! It is
mathematically beautiful and physically natural.

It is important to notice that in general $K_\mu =[\re(K_\mu )+\i\im
(K_\mu )]$ is complex,
\begin{eqnarray} 
&\dsize \i\im(K_\mu )  =\frac{(\bar{R}\gamma _\mu R-\bar{L}\gamma
_\mu
L)(\bar{L}R-\bar{R}L)}{4(\bar{R}L)(\bar{L}R)}
=\fr{(\bar{\Psi}\gamma_\mu\gamma^5\Psi)
(\bar{\Psi}\gamma^5\Psi)}
{(\bar{\Psi}\gamma_\nu\Psi)\eta^{\mu\lambda}(\bar{\Psi}
\gamma_\lambda\Psi)} \,,
\nonumber\\\noa1
&\dsize {\rm Re}\,(K_\mu )  =\frac{(\bar{R}\gamma _\mu R+\bar{L}\gamma _\mu L)(
\bar{R}L+\bar{L}R)}{4(\bar{R}L)(\bar{L}R)}
 =\frac{(\bar{\Psi }\gamma _\mu \Psi )(\bar{\Psi }\Psi )}{(
\bar{\Psi }\gamma _\nu \Psi )\eta ^{\nu \lambda }(\bar{\Psi }
\gamma _\lambda \Psi )}\,.
\end{eqnarray}
Only $\re(K_\mu )$ is associated with the phase angle,
and imaginary part $\im(K_\mu
)$ is associated with the ``scale factor'' $\sigma =\e^{-m\int [\im(K_\mu
)]\d x^\mu }$ of a spinor. (We can find the
similar scale change in the
Weyl's early work.$^{[3]}$) Furthermore
\begin{eqnarray} 
& [\re(K_\mu )] [\im(K^\mu )]=0 \,,
\\
& \bar{\Psi }[\re(K_\mu )]\gamma ^\mu \Psi =\bar{\Psi }\Psi \,,\qquad
\bar{\Psi }[\im(K_\mu )]\gamma ^\mu \Psi =0\,.
\end{eqnarray}
One can realize that cubic-trilinear form is independent of $\im(K_\mu )$.
Thus, if
\begin{equation} 
\bar{\Psi }\,\stackrel{\rm d}{=}\,\Psi ^{*T}\gamma ^0 \,,\qquad
\bar{\bar{\Psi}}_0\,\stackrel{{\rm d}}{=}\,\bar{\Psi}\e^{-\i\int
[(-e)A_\mu -(-m)K_\mu ]\d x^\mu }\,,
\end{equation}
then the Lagrangian
\begin{equation} 
\bar{\bar{\Psi}}_0\i\gamma ^\mu \partial _\mu \Psi _0=\bar{
\Psi }\i\gamma ^\mu (\partial _\mu \Psi -\i eA_\mu )\Psi -m\bar{\Psi }\Psi
\end{equation}
is independent of the ``scale factor'' $\sigma =\e^{-m\int [\im(K_\mu
)]\d x^\mu}$, i.e., is independent of $\im(K_\mu )$.

Sometimes it is prefer to use the Lagrangian formalism as a starting point
in constructing various quantum field theories. The point of Lagrangian
formalism is that it makes it easy to satisfy conformal invariance and
especially to obtain Noether's conserved currents. For this purpose we
introduce null-twistors which satisfy $\bar{\Phi }\Phi =0$. In special
coordinate system they can be written in the following forms
\begin{equation} 
 \Phi _{+}\,\stackrel{{\rm d}}{=}\,\Bigl[1-\i x^\mu \gamma _\mu
\Bigl(\frac{1+\gamma _5}
2\Bigr)\Bigr]R=\exp \Bigl[-\i x^\mu \gamma _\mu
\Bigl(\frac{1+\gamma _5}2\Bigr)\Bigr]R\,,
\end{equation}
or
\begin{equation} 
\Phi _{-}\,\stackrel{{\rm d}}{=}\,\Bigl[1-\i x^\mu \gamma _\mu
\Bigl(\frac{1-\gamma _5}
2\Bigr)\Bigr]L=\exp \Bigl[-\i x^\mu \gamma _\mu
\Bigl(\frac{1-\gamma _5}2\Bigr)\Bigr]L\,.
\end{equation}
The above additional exponential factors for $R$ and $L$ first appeared in
the redefined wavefunction of electron in the work of Dirac,$^{[4]}$
where the wave equation in conformal space was studied. Dirac introduced
physical wavefunction $\phi $, connected with conformal spinor $\psi $ by
$\psi = \{1-\i x^\mu \gamma _\mu [({1+\gamma _5})/2]\}\phi $. From
conformally invariant wave equation for $\psi $, which involves the spin
matrices, he obtained the wave equation for physical spinor $\phi $.
``This equation is equivalent to the usual wave equation for electron,
except for the factor $({1+\gamma _5})/2$, which introduces a degeneracy'', and thus this
equation was rejected by Dirac himself.

The geometry of the above null-spinors is shown clearest in terms of projective
twistor space.$^{[1,5]}$ Considering $\Phi _{+}$ to be fixed and solving
for real solutions $x^\mu \in M^4$ of Eq.~(17) (or Eq.~(18)), it turns
out that a solution exists only if $\bar{\Phi }_{+}\Phi _{+}=0$. These
solutions $x^\mu (\tau )$ (for any fixed $\Phi _{+}$) in real Minkowski
space $M^4$ constitute a null straight line (null geodesic with parameter $
\tau $), and every null straight line in Minkowski space arises in this way.
So a point in Minkowski space is said to be ``incident'' with the null
twistor. This is the so-called standard flat-space twistor correspondence.

The null-twistors are form-invariant under the following conformal
transformations
\begin{eqnarray} 
\tilde{\Phi}_{+}=S^{+}\Phi _{+}=\Bigl[1-\i\tilde{x}^\mu \gamma _\mu
\Bigl(\frac{1+\gamma _5}2\Bigr)\Bigr](S_0R) \,,
\nonumber\\
 \tilde{\Phi}_{-}=S^{-}\Phi _{-}=\Bigl[1-\i\tilde{x}^\mu \gamma _\mu \Bigl(
\frac{1-\gamma _5}2\Bigr)\Bigr](S_0L)\,,
\end{eqnarray}
where the operators which correspond to the above conformal transformations $
S^{\pm }$ are
\begin{equation}
\begin{array}{llll} 
& P_\mu ^{\pm }  &\! =\partial _\mu  &\dsize +\gamma _\mu
\Bigl(\frac{1\pm \gamma _5}2\Bigr) ,
\nonumber\\
& M_{\mu \nu }^{\pm } &\! =(x_\nu \partial _\mu -x_\mu \partial _\nu )
&\dsize -\frac14\sigma _{\mu \nu } \,,
\nonumber\\
& K_\mu ^{\pm } &\! = 2x_\mu x^\lambda \partial _\lambda -x^2\partial _\mu
&\dsize +\gamma _\mu \Bigl(\frac{1\mp \gamma _5}2\Bigr),
\nonumber\\
& D^{\pm }  &\! =-x^\mu \partial _\mu &\dsize \mp \frac 12\gamma _5\,,
\end{array}
\end{equation}
here $\sigma _{\mu \nu }=(\gamma _\mu \gamma _\nu
-\gamma _\nu \gamma _\mu )$.
And the operators which correspond to the induced transformations
$S_0$ are
\begin{equation}
\begin{array}{llll} 
& P_{0\mu } &\! =\partial _\mu\,, & {}
\nonumber\\
& M_{0\mu \nu } &\! =(x_\nu \partial _\mu -x_\mu \partial _\nu ) &
\tsize-\frac14\sigma _{\mu \nu } \,,
\nonumber\\
& K_{0\mu } & \!= 2x_\mu x^\lambda \partial _\lambda -x^2\partial _\mu
& +x^\lambda \gamma _\mu \gamma _\lambda \,,
\nonumber\\
& D_0 &\! =-x^\mu \partial _\mu & \tsize-\frac 12\,.
\end{array}
\end{equation}
Two sets of the above operators satisfy the same commutation relations of
conformal Lie algebra.

Now we are in a position to study conformal property of Dirac field.

\vspace*{0.15cm}
\noindent{\bfit Theorem\ \ } Dirac Lagrangian for massive fermion can be rewritten in the following
``uncharged massless {\bfit bosonic}'' form
\begin{equation}
\begin{array}{ll}  
{\bk L}\sqrt{g}\d^4x &\dsize =[(\partial _\mu \bar{\bar{\Phi}}_{+})g^{\mu \nu
}(\partial _\nu \Phi _{+})+(\partial _\mu \bar{\bar{\Phi}}_{-})g^{\mu \nu
}(\partial _\nu \Phi _{-})]\sqrt{g}\d^4x
\nonumber\\
&\dsize =[\bar{\Psi }\i\gamma _\nu \eta ^{\nu \mu }(\partial _\mu -\i eA_\mu
)\Psi -2m\bar{\Psi }\Psi
 -(\partial _\mu +\i eA_\mu )\bar{\Psi }\i\gamma _\nu \eta ^{\nu \mu
}\Psi ]\d^4x\,,
\end{array}
\end{equation}
here $\Psi =R+L$ and
\begin{eqnarray}
&&\dsize \Phi _{+}  \,\stackrel{{\rm d}}{=}\,\Bigl[1-\i x^\mu \gamma _\mu
\Bigl(\frac{1+\gamma _5}
2\Bigr)\Bigr]\Bigl(\frac 1{\Omega}R_0\Bigr)\,,
 \qquad
R_0 \, \stackrel{{\rm d}}{=}\,R \e^{-\i\int (eA_\mu -mK_\mu )\d x^\mu }\,,
\\\noa4
&&\dsize \Phi _{-} \,\stackrel{{\rm d}}{=}\,\Bigl[1-\i x^\mu \gamma _\mu
\Bigl(\frac{1-\gamma _5}
2\Bigr)\Bigr]\Bigl(\frac 1{\Omega}L_0\Bigr) \,,
\qquad
L_0  \,\stackrel{{\rm d}}{=}\,L \e^{-\i\int (eA_\mu -mK_\mu )\d x^\mu }\,,
\end{eqnarray}
and
\begin{equation}  
g^{\mu \nu }=\Omega ^{-2}\eta ^{\mu \nu } \,,\qquad
\sqrt{g}=\Omega ^4\,.
\end{equation}
The additional factor $1/{\Omega}$ in $\Phi _{\pm }$ means that
we work here with the conformal spinors of degree $-1$.\wen{4}

It is easy to prove that the first ``uncharged massless bosonic'' form
of the above Lagrangian implies the property of invariance under conformal
transformations,$^{[1]}$ here the null-spinors transform in the
way $\tilde{\Phi}_{+}=S^{+}\Phi _{+}$ and $\tilde{\Phi}_{-}
=S^{-}\Phi _{-}$, they are form-invariant.
The representations of $S^{\pm }$ are linear and
independent of $x^\mu $. The second ``original'' form of Lagrangian (22)
is familiar to us and its conformal invariance is by no means apparent, but
due to its equivalence to the first form, it of course must have this
property. We point out that the above Lagrangian is independent of $\Omega $,
this means that it is independent under conformal rescaling.

The induced conformal transformations of (physical components) $R$ and $L$
are defined by
\begin{eqnarray}  
&\dsize \tilde{\Phi}_{+}=S^{+}\Phi _+(x,R,L,\Omega ,\gamma ^\mu )=\Phi _{+}(
\tilde{x},\tilde{R},\tilde{L},\tilde{\Omega },\gamma ^\mu )\,,
\nonumber\\
&\dsize \tilde{\Phi}_{-}=S^{-}\Phi _{-}(x,L,R,\Omega ,\gamma ^\mu )=\Phi _{-}(
\tilde{x},\tilde{L},\tilde{R},\tilde{\Omega },\gamma ^\mu )\,,
\end{eqnarray}
and it is not difficult to prove that
\begin{eqnarray}
&&\dsize \tilde{\Omega }=\Bigm| \frac{\partial \tilde{x}}{\partial x}\Bigm|
^{{-1}/4}\Omega \, , 
 \hskip1.5cm  
\frac{\partial x^\nu }{\partial \tilde{x}^\mu
}[S_0\gamma _\nu S_0^{-1}\lambda ]=\gamma _\mu \,,
\nonumber\\\noa3
&&\dsize \tilde{K}_\mu =\Bigm| \frac{\partial \tilde{x}}{\partial x}\Bigm|
^{1/4}\frac{\partial x^\nu}{\partial\tilde{x}^\mu}K_\nu\,,
\hskip0.6cm 
\d^4\tilde{x}=\Bigm| \frac{\partial \tilde{x}}{\partial x}\Bigm| \d^4x\,,
\\\noa5
&&\dsize \tilde{R}=\lambda ^{-1}S_0 R \e^{\i\int (1-\lambda )[eA_\mu
-mK_\mu ]\d x^\mu}\,, 
\end{eqnarray}
here $\lambda =|{\partial \tilde{x}}/{\partial x}|
^{1/4}$. We see that the induced transformations for (physical components) $
R $ and $L$ are nonlinear. $K_\mu $ is a conformal vector of degree $-1$.
Recall that U(1) gauge field potential $A_\mu $ is a conformal vector of
degree $-1$ too.$^{[4]}$

Now there is no problem to check directly that common Lagrangian for Dirac
field (i.e.\ the second form of Eq.~(22)) is unchanged under the above induced
conformal transformations (27) and (28).

It is important to notice that although we bear firmly in mind that the
expressions of $R_0$ and $L_0$ include the ``scale factor'' $\sigma=\e^{-\int
\im (K_\mu )\d x^\mu }$,  $\im(K_\mu )$ has disappeared from the
cubic-trilinear form (14) and from Lagrangians (16) and (22).

\baselineskip=12pt
\sect{\large\bf III. Non-integrable Exponential Factors}
\vspace*{-0.15cm}
Physical spinor field $\Psi $ which satisfies massive charged Dirac equation
can be rewritten in the following form
\begin{equation}  
\Psi \,\stackrel{{\rm d}}{=}\,\Psi _0 \e^{\i\int (eA_\mu -mK_\mu )\d x^\mu }
 =R_0\e^{\i\int (eA_\mu -mK_\mu )\d x^\mu }+L_0 \e^{\i\int (eA_\mu
-mK_\mu )\d x^\mu}\,,
\end{equation}
where $\Psi _0$ satisfies massless Dirac equation and
\begin{equation}  
K_\mu =\frac{\bar{R}\gamma _\mu R}{2\bar{R}L}+\frac{\bar{L}
\gamma _\mu L}{2\bar{L}R}=\frac{\bar{R}_0\gamma _\mu R_0}{2
\bar{R}_0L_0}+\frac{\bar{L}_0\gamma _\mu L_0}{2\bar{L}_0R_0}\,.
\end{equation}

The connection between non-integrability of phase and electromagnetic
potential $A_\mu $ given here is not new, which is essentially just Weyl's
 principle of gauge invariance$^{[6]}$ in its modern form. It is also
contained in the work of Ivanencko and Fock, who considered a more
general kind of non-integrability based on a general theory of parallel
displacement of half-vector. C.N. Yang  reformulated the concept of a
gauge field in an integral formalism and extended Weyl's idea to the
non-Abelian general case.$^{[7]}$

The non-integrable phases for the wavefunctions were also discussed by
Dirac in 1931,$^{[2]}$ where the problem of monopole was studied. He
emphasized that ``non-integrable phases are perfectly compatible with all
the general principles of quantum mechanics and do not in any way restrict
their physical interpretation''. Dirac conjectured that: ``The change
in phase of a wavefunction round any closed curve may be different for
different wave functions by arbitrary multiplies of $2\pi $''. There
is the famous Dirac relation $eq/(4\pi )=(n/2)$. This means that if the
quantization of electric charge (the universal unit $e$) is accepted,
then the above relation is the law of {\bfit quantization}
of the magnetic pole strength.

Notice, in our case $K^\mu =[\re(K^\mu )+\i\im(K^\mu)]$ is complex.
Thus only $\i\int[eA_\mu -m(\re(K_\mu ))]\d x^\mu $ is associated with the
phase-change, and imaginary part $\int \im(K_\mu )\d x^\mu $ is associated with
the scale-change.

Because of the single-valued nature of the quantum mechanical wavefunction,
we naturally conjecture that:

(i) The {\bfit phase-change} of a wavefunction round any closed
curve must be close to $2n\pi $ where $n$ is some integer, positive or
negative. This integer will be a characteristic of possible singularity.

(ii) The {\bfit scale-change} of a wavefunction round any closed
curve must be close to {\bfit zero}. As mentioned in the previous
section, the Lagrangian is independent of thus scale-change. The
scale-factor can be gauged away by conformal rescaling.

This is a new (very strong) assumption, and cannot be proved, not derived.
It is a conjecture of the overall consistency among all the solutions to the
same equation. The existence of magnetic monopole is an open question yet,
thus in our case it means that the change in $\oint mK_\mu \d x^\mu $ round
any closed curve, with the possibility of there being singularity in $
\re(K_\mu )$, will lead to the law of {\bfit quantization} of physical
constants, including {\bfit mass}.

\sect{\large\bf IV. Illustrative Examples}

\noindent{\bf Example 1} (Plane-wave solution):
The simplest solution of massless Dirac equation is that all the four components of
$\Psi _0$ are constants. Thus the components of complex vector $K_\mu $ are
constants too. So
$$
\Psi =\Psi _0 \e^{-\i m\int K_\mu \d x^\mu }
=\Psi _0 \e^{-\i m(K_\mu x^\mu+a)}
$$
(for simplicity we take $A_\mu =0$). The solutions of this type include
plane-wave solution of free electron which is the most important solution in
the quantum field theory. One realizes that in this case, the positive
time-like vector $mK_\mu $ is nothing but the {\bfit energy-momentum} of the
massive Dirac particle. Notice, there is not singularity in $K_\mu $, thus $
\oint K_\mu \d x^\mu =0$.

\vspace*{0.15cm}
\noindent{\bf Example 2}:
Here we use a special chiral representation of Dirac matrices. Let
\begin{equation}  
\gamma ^0=\left(
\begin{array}{cc}
0 & -I \\
-I & 0
\end{array}
\right) ,\quad  \gamma ^i=\left(
\begin{array}{cc}
0 & \sigma ^i \\
-\sigma ^i & 0
\end{array}
\right) , \quad \gamma ^5=\left(
\begin{array}{cc}
I & 0 \\
0 & -I
\end{array}
\right),
\end{equation}
where $\sigma ^i$ are the Pauli matrices and $\Psi
^T=R_0^T+L_0^T=(R_{01},R_{02},L_{01},L_{02})$. In these notations, for some
physical considerations, we will study the following special solution of static
massless Dirac equation,
\begin{equation}  
\begin{array}{lll}
&\dsize R_{01}  =\frac z{r(x+\i y)}-\frac{\i p}{(x+\i y)}+q \,,
&\dsize\quad
 R_{02}  =\frac 1r \,,
\nonumber\\\noa3
&\dsize L_{01}  =\frac{-1}r  \,, &\quad
\dsize L_{02}  =\frac z{r(x-\i y)}+\frac{\i p}{(x-\i y)}-q\,,
\end{array}
\end{equation}
here $p$ and $q$ are real constants. Thus
\begin{equation}  
\bar{R}_0L_0-\bar{L}_0R_0=0 \,,\qquad
\bar{\Psi }\Psi =\bar{\Psi }_0\Psi _0=\bar{R}_0L_0+\bar{L
}_0R_0=\frac{4q}r\,,
\end{equation}
and
\begin{eqnarray}  
&\dsize K^1=  \frac{xz}{qr(x^2+y^2)}  -\frac{py}{q(x^2+y^2)} \,,\quad
K^2=  \frac{yz}{qr(x^2+y^2)}  +\frac{px}{q(x^2+y^2)}  \,,
\nonumber\\\noa2
&\dsize K^3=  \frac{2z^2+(p^2-1)r^2}{2qr(x^2+y^2)}+\frac q2r\,.
\end{eqnarray}
One realizes that in this special case, $K^\mu $ are real and vary inversely
proportional to the radius $r$, except for the last term in $K^3$. Moreover
this vector is highly singular on the $z$-axis, and includes a ``vortex
part'',
\begin{equation}  
\vec{\bk K}=\frac pq\left[\frac{-y}{(x^2+y^2)}, \frac
x{(x^2+y^2)},  0\right].
\end{equation}
Its form is familiar to us from the Bohm--Aharonov (solenoid) experiment.

Taking integration around closed curve $\Gamma (\tau )$ (in polar
coordinates),
\begin{equation}  
r=r_0 \, , \qquad  \vartheta =\vartheta _0 \,,
\qquad \varphi =\varphi (\tau )\,,
\end{equation}
we have
\begin{equation}  
m\oint_{\Gamma (\tau )}\vec{K}\cdot \d\vec{x}=\Bigl(\frac{mp}
q\Bigr)2\pi =2n\pi\,.
\end{equation}
Because of the single-valued nature of the quantum mechanical wavefunction,
$n$ must be an integer. Its numerical value is a property of singular line.
The $p/q$ is determined by boundary condition. Thus we get additional
quantization condition ${mp}/q=n$.

\vspace*{0.15cm}
\noindent{\bf Example 3}:
Now let us study the following special solution of massive Dirac equation,
\begin{equation}
\begin{array}{lll}
& R_1  =0\,, & \qquad
R_2  =f\e^{\i(wt+vz)} \,,
\nonumber\\\noa1
&\dsize L_1 =\frac{-2\i}mf^{\prime }\e^{\i(wt+vz)} \,, & \qquad
\dsize L_2  =\frac{(w-v)}mf\e^{\i(wt+vz)}\,,
\end{array}
\end{equation}
here $(w^2-v^2)=m^2$ ($m$ is the mass) and $f(u)$ is the function of $
u=(x+\i y)$, $f^{\prime }=(\partial _uf)$. Thus
\begin{equation}  
\bar{R}L-\bar{L}R=  0 \,,\qquad
\bar{R}L+\bar{L}R=  \frac{2(v-w)}m\bar{f}f\,,
\end{equation}
and
\begin{equation}  
 K^1= \frac{-\i}m\Bigl(\frac{f^{\prime }}f-\frac{\bar{f^{\prime }}}{
\bar{f}}\Bigr),
\quad K^2=  \frac 1m\Bigl(\frac{f^{\prime }}f+\frac{\bar{f^{\prime }}}{\bar{f
}}\Bigr),
\quad
 K^3=  \frac 2{m(w-v)}\frac{f^{\prime }}f\frac{\bar{f^{\prime }}}{
\bar{f}}\,.
\end{equation}
The special case is $f=a(x+\i y)^{-\lambda }\e^{\ssc B}$, here $B(u)$
is the function
of $u=(x+\i y)$, and $B^{\prime }=\partial _\mu B$ is nonsingular. Thus
\begin{eqnarray}  
&&\dsize \bar{R}L+\bar{L}R=\frac{2(v-w)a\bar{a}}{m(\bar{u}
u)^\lambda }\e^{\ssc (B+\bar{B})} \,,
\\\noa1
&&\dsize K^1=  \frac{2\lambda y}{m(x^2+y^2)}  +\frac \i m(\bar{B}^{\prime }
-B^{\prime }) \,,
\quad
 K^2=  \frac{-2\lambda x}{m(x^2+y^2)}  +\frac 1m(\bar{B}^{\prime }
+B^{\prime })\,,
\nonumber\\\noa1
&&\dsize K^3=  \frac 2{m(w-v)}\Bigl(\frac \lambda u+B^{\prime }\Bigr)
\Bigl(\frac \lambda {
\bar{u}}+\bar{B^{\prime }}\Bigr).
\end{eqnarray}
One realizes that in this special case, $K^\mu $ are real and highly
singular on the $z$-axis. They include a ``vortex part'', which has
been discussed in the previous example. The additional quantization condition
for the wavefunction of this type is
\begin{equation}  
\lambda =n/2 \,,\qquad n=0,\pm 1,\pm 2,\cdots\,.
\end{equation}

\sect{\large\bf V. Triality and Bosonization of Fermion}
\vspace*{-0.1cm}
One knows that the group Spin$(2n)$ is the double covering group of the
rotation group SO$(2n)$, i.e., this group has two basic half-spinor
(semi-spinor) representations of degree $2^{n-1}$. A particularly
interesting situation which has some relevance in physics is given when $
2n=2^{n-1}$, that is $2n=8$. In this case Spin(8) has just three
irreducible representations of degree 8 which are all real, and three
representation spaces (vector space) $R$, (semi-spinor spaces) $S_{+}$ and $
S_{-}$ are, remarkably, all on an equal footing. There is an extra
automorphism that exchanges representations which would not be related by any
symmetries for other SO$(2n)$ groups. In fact, it turns out that the extra
symmetry, which is known as ``triality'',\wen{8-10}
relates the spinor representations
of SO(8) to the vector representations. The word triality is applied to
the algebraic and geometric aspects of the $\Sigma _3$ symmetry which
Spin(8) has. One needs to know what three objects the symmetric group $
\Sigma _3$ permutes, and the answer is that it permutes representations.

\vspace*{0.2cm}
\noindent {\bfit Theorem\ \ }
(Principle of triality$^{[8,9]}$) There exists an automorphism $J$ of
order 3 of the vector space $A=R\times S_{+}\times S_{-}$ (dimension
$=8+16$)
which has the following properties: $J$ leaves the quadratic form $\Omega $ and
the cubic form $F$ invariant, and $J$ maps $R$ onto $S_{+}$, $S_{+}$
onto $S_{-}$, and $S_{-}$ onto $R$.

The law of composition in the algebra $A$ is defined in terms of the
quadratic forms $\Omega $ and cubic form $F$ only. And it is clear that any
automorphism of the vector space $A$ which leaves these forms invariant is
an automorphism of the algebra $A$.

However physicists work in 4-dimensional Minkovski space-time $M^{1+3}$ with
signature ($+,-,-,-$). It is important to notice that the minimum dimension
of gamma matrices, and thus the number of complex spinor components, depends
on both dimension of space-time and {\bfit signature} of metric.
In 4-Lorentzian dimensions, gamma matrices are (at least) $4\times 4$,
and thus the number of {\bfit complex} spinor components is four too. The
quadratic-two-form in vector space is defined by means of 
$\eta _{\mu\nu}={\rm diag}\,(+1,-1,-1,-1)$, 
and the quadratic-two-form in spinor space is
defined by means of Dirac conjugate spinors ($\bar{\psi }\psi$, here $
\bar{\psi }=\psi^{*T}\gamma ^0$). We will prove that there exists a
double covering vector representation of Dirac spinors. In other words there
exists an automorphism $J$ of order 3 in $A=M^{1+3}\times
S_{1}^4\times S_{2}^4$ (vector space and two half-spinor spaces), which
leaves quadratic form and cubic form invariant up to the sign. (In Minkowski
space we are faced with spacelike and timelike vectors, thus sometimes it
is convenient to use term {\bfit pseudoscalar},
here the prefix ``{\bfit pseudo}'' refers  to automorphism~$J$.)

The situation is differ from the case of SO(8), and we
propose to denominate this symmetry as a ``{\bfit ding}'' symmetry.
(Chinese {\bfit ding} is an ancient vessel which has two loop
handles and three legs.)

Let us first introduce trinomial unit-basis $(j^\mu ,f^\alpha ,\varphi
^\beta )$, where the basic unit vector $j^\mu $ and two basic unit spinors $%
f^\alpha $, $\varphi ^\beta $ are normalized such that
\begin{eqnarray}
 && j^\mu \eta _{\mu \nu }j^\nu =-1\,, \hskip2cm \bar{f}f=-1 \,,
\hskip2cm \bar{\varphi }\varphi =1 \,,
\nonumber\\
&& j^\mu =-\bar{\varphi }\i\gamma^\mu f=\bar{f}\i\gamma ^\mu \varphi \,,
 \hskip0.6cm  f=j_\mu \i\gamma ^\mu \varphi, 
\hskip1.65cm \varphi =j_\mu \i\gamma ^\mu f\,,
\\\noa1
&& \bar{\varphi }f=\bar{f}\varphi =0 \,,
\hskip2.1cm  
j_\mu (\bar{\varphi }
\gamma ^\mu \varphi )=j_\mu (\bar{f}\gamma ^\mu f)=0\,, 
\\
&&
j_\mu (\bar{\varphi }\i\gamma^\mu f)=-j_\nu (\bar{f}\i\gamma ^\nu \varphi )
=1\,.
\end{eqnarray}
In addition, we can introduce another unit vector $k^\mu $ which is
determined by the above trinomial unit-basis and will play very important role
in our theory. That is
\begin{equation}  
k^\mu \,\stackrel{{\rm d}}{=}\,\bar{\varphi }\gamma ^\mu \varphi =\bar{f}
\gamma ^\mu f \, , \qquad  k^\mu k_\mu =1 \, , \qquad k^\mu j_\mu =0\,.
\end{equation}
Furthermore we need the following algebraic properties
$$\displaylines{\hskip3cm
\arraycolsep1pt 
\begin{array}{lll}
 \tsize \frac 12(\bar{\varphi }\gamma ^\lambda \gamma ^\nu \gamma ^\mu f+
\bar{f}\gamma ^\mu \gamma ^\nu \gamma ^\lambda \varphi ) && 
=\epsilon^{\mu \nu \lambda \rho }k_\rho \,\stackrel{{\rm d}}{=}\,
 \epsilon ^{\mu \nu \lambda} \,,
\nonumber\\
\tsize\frac 12\bar{\varphi}(\gamma ^\mu \gamma ^\nu \gamma ^\lambda +\gamma
^\lambda \gamma ^\nu \gamma ^\mu )\varphi  && 
= t^{\mu \nu \lambda \rho
}k_\rho  \,\stackrel{{\rm d}}{=}\, t^{\mu \nu \lambda } \,,
\nonumber\\
 \tsize\frac 12\bar{f}(\gamma ^\mu \gamma ^\nu \gamma ^\lambda +\gamma
^\lambda \gamma ^\nu \gamma ^\mu )f && 
= t^{\mu \nu \lambda \rho }k_\rho
\,\stackrel{{\rm d}}{=}\, t^{\mu \nu \lambda } \,,
\nonumber\\
 \tsize\frac 12\bar{\varphi }(\gamma ^\mu \gamma ^\nu \gamma ^\lambda -\gamma
^\lambda \gamma ^\nu \gamma ^\mu )\varphi
&& 
=\i\epsilon ^{\mu \nu \lambda
\rho }j_\rho  \,\stackrel{{\rm d}}{=}\,
\i\check{\epsilon}^{\mu \nu \lambda } \,,
\nonumber\\
 \tsize\frac 12\bar{f}(\gamma ^\mu \gamma ^\nu \gamma ^\lambda -\gamma
^\lambda \gamma ^\nu \gamma ^\mu )f  && 
=\i\epsilon ^{\mu \nu \lambda \rho
}j_\rho  \,\stackrel{{\rm d}}{=}\,\i\check{\epsilon}^{\mu \nu \lambda } \,,
\nonumber\\
\tsize\frac 12(\bar{\varphi }\gamma ^\lambda \gamma ^\nu \gamma ^\mu f-
\bar{f}\gamma ^\mu \gamma ^\nu \gamma ^\lambda \varphi )
 &&  = \i t^{\mu \nu
\lambda \rho }j_\rho \,\stackrel{{\rm d}}{=}\,
\i\check{t}^{\mu \nu \lambda }\,,
\end{array}\hfill\cr\noalign{\vskip-15pt}
\hfill(48)\cr}
$$
$$\displaylines{\hskip3.1cm
\bar{\varphi }\gamma ^\mu \gamma ^\nu \varphi   =  -\bar{f}
\gamma ^\mu \gamma ^\nu f  =  \eta ^{\mu \nu }+\i\epsilon ^{\mu \nu \lambda
\rho }k_\lambda j_\rho  \,, 
\hfill\cr\hskip3.1cm
\bar{\varphi }\gamma ^\mu \gamma ^\nu f  =  \bar{f}\gamma ^\mu
\gamma ^\nu \varphi   =  \i(k^\mu j^\nu -j^\mu k^\nu )\,,
\hfill(49)\cr}
$$
here $\epsilon ^{\mu \nu \lambda \rho }$ is the Levi--Civita symbol with the
definition $\epsilon ^{0123}=1$, and
\setcounter{equation}{49}
\begin{eqnarray}  
&\tsize \epsilon ^{\mu \nu \lambda \rho }=\frac \i4\tr(\gamma ^5\gamma ^\mu \gamma
^\nu \gamma ^\lambda \gamma ^\rho ) \,,
\nonumber\\
&\tsize t^{\mu \nu \lambda \rho } \,\stackrel{{\rm d}}{=}\,
\frac 14\tr(\gamma ^\mu \gamma ^\nu
\gamma ^\lambda \gamma ^\rho )=(\eta^{\mu \nu }\eta^{\lambda \rho }
+\eta^{\mu \rho}\eta^{\nu \lambda }-\eta^{\mu \lambda}\eta^{\nu \rho })\,.
\end{eqnarray}
These objects provide the required algebraic multiplication rule on the
algebra $A$, that are needed. Let
\begin{eqnarray}  
& \gamma ^{\mu \nu \lambda }\,\stackrel{{\rm d}}{=}\,\epsilon ^{\mu \nu \lambda
}+\i t^{\mu \nu \lambda }=(-\gamma^{\lambda\nu\mu})^* \,,
\nonumber\\
&  \check{\gamma}^{\mu \nu \lambda }
\,\stackrel{{\rm d}}{=}\,\check{\epsilon}^{\mu \nu \lambda }
+\i\check{t}^{\mu \nu \lambda }=(-\gamma^{\lambda\nu\mu})^*\,,
\end{eqnarray}
then
\begin{equation}
\begin{array}{lll}
& \gamma ^{\mu \nu \sigma }\eta_{\sigma \delta }\gamma ^{\lambda \rho \delta
}+\gamma ^{\mu \rho \sigma }\eta_{\sigma \delta }\gamma ^{\lambda \nu \delta
}=& \,-\,2\eta^{\mu \lambda }\eta^{\nu \rho } \,,
\nonumber\\
& \check{\gamma}^{\mu \nu \sigma }\eta_{\sigma \delta }\check{\gamma}^{\lambda
\rho \delta }+\check{\gamma}^{\mu \rho \sigma }\eta_{\sigma \delta }\check{
\gamma}^{\lambda \nu \delta }=& 2\eta^{\mu \lambda }\eta^{\nu \rho }\,.
\end{array}
\end{equation}

The most important to us is: the trinomial unit-basis $(\varphi ^\beta
,f^\alpha ,j^\mu )$ which satisfies the above conditions exists (It
can be  easily verified from the special case (56)).

For better understanding geometrical and physical interpretation of complex
vector $K^\mu $ introduced in the previous section, it is convenient
passing from trinomial (tripartite) unit-basis $(\varphi _1^\beta ,f_2^\alpha
,j_3^\mu )$ to the null-basis $(r^\alpha ,l^\beta ,k_{\pm }^\mu )$, where
the normalized right-handed spinor $r$ and left-handed spinor $l$ are
determined in the following way
\begin{equation}
\begin{array}{llll}
& \bar{r}l=\bar{l}r=2 \, , & \qquad f=(\i/2)(r-l) \,,
& \qquad\tsize \varphi =\frac 12(r+l) \,,
\nonumber\\
&\tsize k_{\pm }^\mu =\frac 12(k^\mu \pm j^\mu ) \, ,
& \qquad k_{\pm }^\mu \eta _{\mu \nu
}k_{\pm }^\nu =0 \, ,
 &\tsize \qquad k_{\pm }^\mu \eta _{\mu \nu }k_{\mp }^\nu =\frac 12\,.
\end{array}
\end{equation}

It is easy to prove that
\begin{equation}  
 k_\mu \gamma ^\mu r=k_\mu ^{-}\gamma ^\mu r=l \,, \qquad
k_\mu \gamma ^\mu l=k_\mu ^{+}\gamma ^\mu l=r \,,
\end{equation}
and it is nothing but the special case of Eq.~(3).

We know that Dirac spinor includes four complex components, or equivalently
$4+4=8$ independent real components. Any Dirac spinor can be decomposed in the
sum of two `half-spinors' $\Psi =\Psi _1+\Psi _2$ which can be constructed
by means of $f$ and $\varphi $, i.e.,
\begin{equation}  
\Psi _1=B^\mu \eta _{\mu \nu }\i\gamma ^\nu f \,,
\qquad \Psi _2=N^\mu \eta _{\mu\nu }\i\gamma ^\nu \varphi\,,
\end{equation}
here $B^\mu $ and $N^\mu $ are real vectors.

In order to easily understand our idea, it is convenient to work in the
special coordinate system such that
\begin{equation}  
[ f^\alpha ]^T=(0,0,\i,0) \,,\quad
[\varphi ^\beta ]^T=(1,0,0,0) \,, \quad j^\mu =(0,0,0,1) \,,\quad
k^\mu =(1,0,0,0) \,.
\end{equation}
(Here we use Dirac representation of gamma matrices.) In this special case
\begin{equation}  
\Psi =\Psi _1(B)+\Psi _2(N)=\left(
\begin{array}{c}
B^3+\i N^0 \\ 
B^1+\i B^2 \\
B^0+\i N^3 \\
-N^1+\i N^2
\end{array}
\right) .
\end{equation}

Let $V^\mu \in M^{1+3}$ be the vector in Minkovski space, $\Psi _1(B)\in S_1$
and $\Psi _2(N)\in S_2$ are two half-spinors referred to trinomial
unit-basis defined as in the above. The quadratic two-forms and the cubic
trilinear-form are well defined, i.e.,
\begin{eqnarray}  
& V^\mu \eta _{\mu \nu }V^\nu =V_\nu V^\nu \,, \quad \bar{\Psi }_1\Psi
_1=-B_\nu B^\nu \,, \quad \bar{\Psi }_2\Psi _2=N_\mu N^\mu\,,
\\\noa0
& V^\mu \eta _{\mu \nu }(\bar{\Psi }_1\gamma ^\mu \Psi _2+\bar{\Psi }
_2\gamma ^\mu \Psi _1)=2(\epsilon _{\mu \nu \lambda \rho }k^\mu )V^\nu
B^\lambda N^\rho\,.
\end{eqnarray}
Furthermore
\begin{equation}  
\tsize B^\mu =\frac 12(\bar{f}\i\gamma ^\mu \Psi _1-\bar{\Psi }_1\i\gamma
^\mu f) \,, \qquad
N^\mu =\frac 12(\bar{\Psi }_2\i\gamma ^\mu \varphi -\bar{\varphi }
\i\gamma ^\mu \Psi _2)
\end{equation}
define the vector representations of half-spinors $\Psi _1$ and $\Psi _2$. One
realizes that there exists a ``ding'' automorphism $J$ of order 3 in $
M^{1+3}\times S_1\times S_2$, which leaves quadratic two-forms and the above
cubic trilinear-form invariant up to the sign. $J$ maps $M$ onto $S_1$, $S_1$
onto $S_2$, and $S_2$ onto $M$ (Or equivalently $V\rightarrow B\rightarrow
N\rightarrow V$).

The most interesting for physicists is that: by passing from ordinary spinor
representation to the vector representation, one can express Dirac
Lagrangian in the bosonic form
\begin{equation}  
\begin{array}{ll}
&\quad\tsize \frac 12[\bar{\Psi }\i\gamma ^\mu (\partial _\mu -\i eA_\mu )\Psi
-((\partial _\mu +\i eA_\mu )\bar{\Psi })\i\gamma ^\mu \Psi ]-m\bar{
\Psi }\Psi
\nonumber\\
&\dsize  =  -[(B_\nu \partial _\lambda B_\rho +N_\nu \partial _\lambda N_\rho )%
\check{\epsilon}^{\nu \lambda \rho }-(B_\nu \partial _\lambda N_\rho -N_\rho
\partial _\lambda B_\nu )\check{t}^{\nu \lambda \rho }]
\nonumber\\
&\dsize\hphantom{= }\, +\; m(B_\mu B^\mu -N_\nu N^\nu )+eA_\mu [(B_\nu B_\lambda +N_\nu N_\lambda
)t^{\nu \mu \lambda }+2B_\nu N_\lambda \epsilon ^{\nu \mu \lambda }]\,,
\end{array}
\end{equation}
where $e$ and $m$ are the charge and mass of the physical particle. The
Lagrangian is invariant under U(1) gauge transformation
\begin{eqnarray}  
& \widetilde{\Psi }=\Psi  \e^{\i\alpha }=\Psi (\cos \alpha +\i\sin \alpha )\,,
\qquad \widetilde{A_\mu }=A_\mu +\partial _\mu \alpha\,,
\nonumber\\
& \widetilde{B_\mu }=B_\mu \cos \alpha -[\check{\epsilon}_{\mu \nu
\lambda }k^\lambda B^\nu -(k_\mu j_\nu -k_\nu j_\mu )N^\nu ]\sin \alpha \,,
\nonumber\\
& \widetilde{N_\mu }=N_\mu \cos \alpha -[\check{\epsilon}_{\mu \nu
\lambda }k^\lambda N^\nu +(k_\mu j_\nu -k_\nu j_\mu )B^\nu]\sin \alpha \,,
\end{eqnarray}
where $\e^{\i\alpha }=\cos \alpha +\i\sin \alpha $.

The corresponding massive Dirac equation (which has eight independent equations)
in the vector representation takes the form of
\begin{equation}  
\left\{
\begin{array}{l}
 \check{\epsilon}^{\mu \nu \lambda }\partial _\nu B_\lambda -\check{t}^{\mu
\nu \lambda }\partial _\nu N_\lambda -eA_\nu (t^{\mu \nu \lambda }B_\lambda
+\epsilon ^{\mu \nu \lambda }N_\lambda )-mB^\mu =0 \,,
\\
\check{\epsilon}^{\mu \nu \lambda }\partial _\nu N_\lambda +\check{t}^{\mu
\nu \lambda }\partial _\nu B_\lambda -eA_\nu (t^{\mu \nu \lambda }N_\lambda
-\epsilon ^{\mu \nu \lambda }B_\lambda )+mN^\mu =0\,,
\end{array}\right.
\end{equation}
or equivalently
\begin{equation}  
\left\{
\begin{array}{l}
\partial _\mu N^\mu -[mB^\mu +eA_\nu (B_\lambda t^{\mu \nu \lambda
}+N_\lambda \epsilon ^{\mu \nu \lambda })]j_\mu =0 \,,
\\\noa0
 \partial _\mu B^\mu -[mN^\mu -eA_\nu (N_\lambda t^{\mu \nu \lambda
}-B_\lambda \epsilon ^{\mu \nu \lambda })]j_\mu =0 \,,
\\\noa0
 \nabla _\mu N_\nu -\nabla _\nu N_\mu =\frac 12\epsilon _{\mu \nu \lambda
\rho }(\nabla ^\lambda B^\rho -\nabla ^\rho B^\lambda )\,,
\end{array}\right.
\end{equation}
where
\begin{eqnarray}  
&\tsize \nabla _\mu N_\nu \,\stackrel{{\rm d}}{=}\,\partial _\mu N_\nu
+(m/2)\check{%
\epsilon}_{\mu \nu \rho }N^\rho +e\bigl(j_\mu \eta _{\nu \sigma }\epsilon
^{\sigma \lambda \rho }-\frac 12\check{\epsilon}_{\mu \nu \sigma }t^{\sigma
\lambda \rho }\bigr)A_\lambda N_\rho  \,,
\nonumber\\ 
&\tsize \nabla _\mu B_\nu \,\stackrel{{\rm d}}{=}\,\partial _\mu B_\nu -
(m/2)\check{%
\epsilon}_{\mu \nu \rho }B^\rho +e\bigl(j_\mu \eta _{\nu \sigma }\epsilon
^{\sigma \lambda \rho }-\frac 12\check{\epsilon}_{\mu \nu \sigma }t^{\sigma
\lambda \rho }\bigr)A_\lambda B_\rho \,.
\end{eqnarray}

In above sense the half-spinor of the first type is dual to the half-spinor
of the second type.  We can define ``dual'' transformation $B\to N$, $N\to-B$,
$m\to-m$ which leaves quadratic form $m\bar{\Psi}\Psi=m(B_\mu
B^\mu-N_\mu N^\mu)$, cubic form $A_\mu B_\nu N_\lambda
\epsilon^{\mu\nu\lambda}$ and thus the above Lagrangian and Dirac
equation invariant. (It looked like electro-magnetic duality which
accompanied by $e\to q$ and $q\to -e$.)

If we pass from trinomial (tripartite) unit-basis $(f^\alpha ,\varphi ^\beta
,j^\mu )$ to the null-basis $(r^\alpha ,l^\beta ,k_{\pm }^\mu)$,
then the spinor can be decomposed in the sum of right-handed
and left-handed spinors $\Psi =R+L$, here
\begin{equation}  
\tsize  R=\frac 12G^\mu \eta _{\mu \nu }\gamma ^\nu l \, , \quad
 L=-\frac{1}2G^{*\mu}\eta _{\mu \nu }\gamma ^\nu r\,,
\quad
 G^\mu \,\stackrel{{\rm d}}{=}\, (B^\mu +\i N^\mu )\,.
\end{equation}
In these notations Dirac Lagrangian takes the form of
\begin{equation}  
L=(\partial_\mu G^*_\nu)\check{\gamma}^{\nu\mu\lambda}G_\lambda
-G^*_\nu\check{\gamma}^{\nu\mu\lambda}(\partial_\mu
G_\lambda)
 -2\i e A_\mu G^*_\nu \gamma^{\nu\mu\lambda}
G_\lambda  +m(G^*_\nu G^{*\nu}+G_\nu G^\nu)\,.
\end{equation}
The ordinary Dirac equation for massive fermion takes the form of
\begin{equation}  
\check{\gamma}^{\mu\nu\lambda}\partial_\nu G_\lambda
+\i e \gamma^{\mu\nu\lambda}A_\nu G_\lambda-mG^*_\mu=0\,.
\end{equation}
It can be rewritten in another self-dual form
\begin{equation}  
\partial _\mu G^\mu -\i mj_\mu G^{*\mu }=0 \,,\qquad
G_{\mu \nu }=(\i/2)\ \epsilon _{\mu \nu \lambda \rho }G^{\lambda \rho }\,,
\end{equation}
here for simplisity we take $A_\mu =0$ and
\begin{equation}  
G_{\mu \nu }\,\stackrel{{\rm d}}{=}\,[(\partial _\mu G_\nu +\i mj_\mu G_\nu
^{*})-(\partial _\nu G_\mu +\i mj_\nu G_\mu ^{*})]\,.
\end{equation}
Let us define operator $\nabla _\mu \,\stackrel{{\rm d}}{=}\,(\partial _\mu
+\i mj_\mu {\cal C}^{*})$, such that
\begin{equation}  
\nabla _\mu G_{\nu \lambda }\,\stackrel{{\rm d}}{=}\,\partial _\mu G_{\nu \lambda
}+\i mj_\mu G_{\nu \lambda }^{*} \,,
\end{equation}
where ${\cal C}^{*}$ is the operator of complex conjugation: $
{\cal C}^{*}\Phi =\Phi ^{*}$. In this notation the identities
\begin{equation}  
\nabla _\mu G_{\nu \lambda }+\nabla _\lambda G_{\mu \nu }+\nabla _\nu
G_{\lambda \mu }=0
\end{equation}
look like the Bianchi identities, and
\begin{equation}  
G_{\mu \nu }\epsilon ^{\mu \nu \lambda \rho }G_{\lambda \rho }+G_{\mu \nu
}^{*}\epsilon ^{\mu \nu \lambda \rho }G_{\lambda \rho }^{*}
=  2\partial _\mu [\epsilon ^{\mu \nu \lambda \rho }(G_\nu G_{\lambda \rho
}+G_\nu ^{*}G_{\lambda \rho }^{*})]
\end{equation}
look like the Chern--Pontryagin density and a total derivative of the
Chern--Simons density. Thus the Dirac Lagrangian can be modified by the
additional total derivative of the Chern--Simons density introduced in the
above.

The real part $\re(K^\mu )$ and imaginary part $\im(K^\mu )$ which
were
introduced in the previous sections take the form of
\begin{eqnarray}  
& {\rm Re}\,(K^\mu )=-\pi^\mu (k)[G^{*\lambda }G_\lambda ^{*}+G^\lambda G_\lambda
]/[2(G^\nu G_\nu )(G^{*\lambda }G_\lambda ^{*})]\,,
\nonumber\\
& \i\im(K^\mu )=-\pi _5^\mu (j)[G^{*\lambda }G_\lambda ^{*}-G^\nu G_\nu
]/[2(G^\nu G_\nu )(G^{*\lambda }G_\lambda ^{*})]\,,
\end{eqnarray}
here
\begin{eqnarray}  
& \pi^\mu (k)  = \bar{\Psi }
\gamma ^\mu \Psi =(t^{\nu \mu \lambda }+\i\epsilon ^{\nu \mu \lambda
})G_\nu G_\lambda ^{*}  \,,
\nonumber\\ 
& \pi _{5}^\mu (j)  = \bar{\Psi}\gamma^\mu \gamma^5\Psi
=(\check{t}^{\nu \mu \lambda }+\i\check{\epsilon}^{\nu \mu
\lambda })G_\nu G_\lambda ^{*} \,.
\end{eqnarray}
It means that $\re(K^\mu )$ is associated with unit vector $k^\mu $ and $
\im(K^\mu )$ is associated with unit vector $j^\mu $. Furthermore
\begin{equation}  
\re(K^\mu )\eta _{\mu \nu }\im(K^\nu )=k^\mu \eta _{\mu \nu }j^\nu =0\,.
\end{equation}

\vfill
\end{document}